\documentclass[prb,amsmath,amssymb]{revtex4}

\usepackage{graphicx}

\begin{document}



\title{What Quantity is Measured in an Excess Noise Experiment?}

\author{U. Gavish, Y. Imry, Y. Levinson}
\affiliation{ Weizmann Institute of Science \\ 76100 Rehovot,
Israel}
\author{B. Yurke}
\affiliation{Bell Labs, 1C363\\
Murray Hill, NJ 07974}

\begin{abstract}
Consider a measurement in which the current coming out of a
mesoscopic sample is filtered around a given frequency, amplified,
measured and squared. Then this process is repeated many times and
the results are averaged. Often, two such measurements are
performed on the same system in and out of equilibrium (the
nonequilibrium state can be obtained by a variety of methods,
e.g., by applying a DC voltage or electromagnetic radiation to the
sample). The excess noise is defined as the difference in the
noise between these two measurements. We find that this excess
noise is given by the excess of the non-symmetrized power-spectrum
of the current-noise. This result holds for a rather general class
of experimental setups.
\end{abstract}

\maketitle
\section{Introduction}
\label{sec intro}
Consider a measurement in which the current coming out of a mesoscopic
sample is filtered around a frequency $\Omega>0$, then amplified,
measured,  squared, then this process is repeated many times and the results are averaged.
 The final result of such a procedure is
called the \emph{current fluctuations}, the \textit{power spectrum},  or
 the \textit{current noise}, at frequency $\Omega$.

Often, two such measurements are performed on the same system: the
first while it is driven out of equilibrium (e.g., by applying a
DC voltage \cite{Heiblum} or electromagnetic radiation
\cite{Prober photons,Glattli photons} to it) and the second in
equilibrium (the voltage source is turned off, i.e., the power
supply becomes a short). The \textit{excess noise} is defined as
the difference in the noise between the first and the second
measurement. The present work analyzes what quantity one should
calculate in order to predict excess noise.

In the rest of the introduction some basic concepts in amplification theory are
introduced. In Sec. \ref{sec noise measur proc} the measurement procedure is defined.
In Sec. \ref{sec which quant is meas Class} we analyze the classical case and in
sections \ref{sec which quant is meas Quant} and \ref{sec excess quan noise meas}
 the quantum one - when $\hbar\Omega$ is comparable with or larger than the temperature,
 the voltage, or the RF radiation frequency applied to the sample. Finally, a possible verification of the results is presented.

Our main result can be stated as follows. The result of a noise measurement depends on the particular
instrumentation used in the setup - what type of amplifier and detector are used, what are their temperatures, etc.
It is generally, and usually, neither the fourier transform of the current correlator in the sample,
nor its symmetrized version. However, the \emph{excess} noise is instrumentation-independent,
and is equal to the amplifier-gain squared times the  difference in the fourier transform of
 the current correlator, in and out of equilibrium.
This difference has a clear physical meaning: it is the difference in the power emitted \emph{from}
the sample and \emph{into} the filter, in and out of equilibrium.

\subsection{Cosine and sine components of time dependent functions}
\label{sec cos & sin comp}

A real (and well behaved) function, $I(t)$, can be fourier-represented as:
\begin{eqnarray}\label{FT It}
I(t)=\frac{1}{2\pi}\int_{0}^{\infty}d\omega
\left[I(\omega)e^{-i\omega t} +I^{\ast}(\omega)e^{i\omega t} \right].
\end{eqnarray}
where $
I(\omega)=\int_{-\infty}^{\infty}dt
I(t)e^{i\omega t}.$
If $I(\omega)$ is negligible outside a narrow bandwidth $\Delta_f$ around a center frequency
 $\Omega >0$, $\Delta_f \ll \Omega$, then it is useful to write $I(t)$
in the form \cite{Yurke denker}:
\begin{eqnarray}\label{It cos + sin }
I(t)=I_c(t)\cos \Omega t+I_s(t)\sin \Omega t,
\end{eqnarray}
where are $I_c(t)$ and $I_s(t)$ are real and slowly varying - they have fourier-components only at frequencies
smaller than $\Delta_f$.
$I_c(t)$ and $I_s(t)$ are called the \textit{cosine} and \textit{sine} components of $I(t)$.
Defining the time average of $f(t)$ as:
\begin{eqnarray}\label{time average defined}
\overline{f(t)}\equiv \frac{1}{T_0}\lim_{T_0 \rightarrow \infty}\int_{t}^{t+T_0}dt' f(t'),
\end{eqnarray}
then (if $\Delta_f \ll \Omega$) Eq. (\ref{It cos + sin }) implies:
$\overline{I^2(t)}= \frac{1}{2}\overline{I_c^2(t)}+\frac{1}{2}\overline{I_s^2(t)}.$
In a \textit{stationary state}, the choice of the
origin of time does not affect average quantities and therefore
using  and Eq. (\ref{It cos + sin })  for $t=0$ and $t=\pi/(2\Omega)$
one gets:
\begin{eqnarray}\label{time average I2 Ic2 Is2 statinary st}
\overline{I^2(t)}= \overline{I_c^2(t)}=\overline{I_s^2(t)}.
\end{eqnarray}
Moreover, in a stationary state time averaging may be replaced by averaging over realizations.
We choose these realizations as repeated measurement on the same system
at $N\gg 1$ different times, $t_n$, which are distributed over a whole time interval, $T_0$,
 which is longer than other time scales in the systems:
 $\hbar /eV$, $\hbar/ (k_BT)$, $\Omega^{-1}$, and $\Delta_f^{-1}$ ( in practice, usually the largest time is
  $\Delta_f^{-1}$ and therefore it is enough to require that the measurements are distributed over a time interval longer than it.).
Thus, we can replace the time average: $\overline{f(t)}\rightarrow \langle f(t) \rangle = \frac{1}{N}\sum_{n=1}^{N}f(t_n).$
\subsection{Linear amplifiers}
\label{sec lin ampl}

Below, we shall consider only \emph{linear} amplifiers.
An amplifier is  linear if the current coming out of it,
$I_a(t)$, is related to the one entering it,  $I_f(t),$ by:
\begin{eqnarray}
\label{Lin ampl defined}
I_a(t)=G_1 I_{f,c}(t)\cos\Omega t
+G_2I_{f,s}(t)\sin \Omega t,
\end{eqnarray}
where $I_{f,c}(t)$ and $I_{f,s}(t)$ are the cosine and sine components of $I_f(t).$
and  where at least one of the numbers $G_1$ and $G_2$ is large.
If $G_1= G_2$ the amplifier is
called \textit{phase-insensitive}.
A phase  insensitive amplifier does what
one would naively expect from an amplifier - it just multiplies the incoming signal by a
 large number.
 If $G_1\neq G_2$  the amplifier is called \textit{phase sensitive} and it affects
the two components of the incoming signal differently. An important special case
is when $G_i\gg  1 \gg G_j,$ $i\neq j,$
where the amplifier amplifies only one of the components
while disposing of the other.

Phase sensitive and insensitive
amplifiers have similar classical behaviors, but may have different technical advantages.
However,
 when quantum effects are important they differ fundamentally due to the
 limitations Heisenberg principle puts on them - see Ref. [4]
 and the Appendix.

 In noise measurement in mesoscopic systems phase insensitive amplifiers are commonly used
  (e.g., in setups that include
 a field effect transistor) but we consider below also the phase sensitive case
 such as the degenerate parametric amplifiers \cite{Yurke denker}
because their quite developed technology
  (that was used, for example, in order
 to enable sensitive detection in experimental gravitational physics \cite{garavitation})
 seems to be less familiar  in the  mesoscopic community   and also
 in order to demonstrate the universality of our result.

\begin{figure}
     \begin{center}
         \includegraphics[height=2.35in,width=4.75in,angle=0]{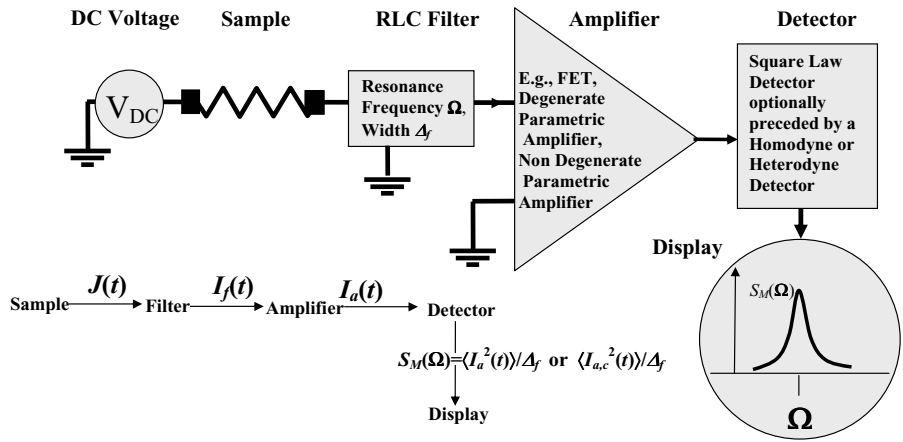}
         \caption{Noise Measurement Procedure}
     \end{center}
\vspace{-0.65cm}
 \end{figure}

\section{Noise and excess noise measurement procedures}
\label{sec noise measur proc}

\subsection{Noise measurement}
\label{sec noise meas proc 1} In a typical noise measurement
(shown in Fig. 1) at frequency $\Omega$
 the current $J(t)$ that flows out of the sample (which is assumed to be in
 a stationary state, but not necessarily in equilibrium,)
 is filtered with an RLC circuit around a resonance frequency $\Omega =(LC)^{-\frac{1}{2}}$.
 The current coming out of the filter, $I_f(t)$, is amplified,
 measured at some arbitrary time point, $t$, and the result of this measurement,
 $I_a(t)$, is squared (say, by a square law detector).
This  measurement is then repeated at  $N\gg 1$ different times, $\{t_n\}$, and the results
are averaged and divided by the filter bandwidth, $\Delta_f$,
 giving a number which we shall denote by $S_M(\Omega)$,
\begin{eqnarray}\label{SMw defined}
S_M(\Omega)= \frac{1}{N\Delta_f}\sum_{n=1}^N I^2_a(t_n).
\end{eqnarray}
Finally, a voltage proportional to $S_M(\omega)$ is sent out to drive a display.

The explanation for why the above measurement is, at least in the classical case,
 a measurement of the current spectrum is given in the next section. Meanwhile we make the
following comments:

1. The above procedure is not a measurement of a function
of time. It is merely a series of \textit{independent} samplings of a stationary process.
Though many setups perform these samplings at times which are separated by a
\textit{constant} time interval, we stress that this is
\textit{not} essential (and actually may create confusion):
 the result will
be the same even if the measurement is performed at \textit{random
times} as explained below Eq. (\ref{time average I2 Ic2 Is2 statinary st})).

2. In some setups it is convenient to convert the signal to a low-frequency one
and measure only the sine or cosine components, for example, by mixing the signal with a local oscillator
(as is done in Heterodyne and Homodyne detection),
 that is, multiplying it by a pure sine or a cosine
and averaging the result over time before squaring it. Such a mixing may introduce
additional noise that should be taken into account. Numerical factors that may multiply
the signal as a result of such a procedure are then cancelled by, e.g.,
calibration of the setup with respect
to a source of noise with a known power spectrum such as a resistor in thermal equilibrium.

If the system, the filter and the amplifier are all in a stationary state, then according
to Eq. (\ref{time average I2 Ic2 Is2 statinary st})
such a procedure yields the same $S_M(\Omega)$ as in the case of measuring and squaring the whole signal.

We shall always assume that the sample and the filter are in a stationary state, but
we shall not necessarily assume that this is the case for the amplifier. It is typically the case for semiconductor
amplifiers such as the field effect transistors used in noise measurement in mesoscopic
 systems but it is  \textit{not}  the case in several types of parametric amplifiers \cite{Yurke denker}.

\subsection{Excess noise measurement}
\label{sec noise meas proc}
In an excess noise measurement one subtracts the noise measured when the system is in equilibrium
from that which is measured when the same system is driven out of equilibrium:
\begin{eqnarray}\label{excess noise defined}
S_{M,excess}(\Omega)= S_{M,noneq}(\Omega) - S_{M,eq}(\Omega).
\end{eqnarray}
The excess noise is useful when one is interested in
looking into the changes in the system which are due to driving it out of equilibrium.
It is also useful when a particular setup (amplifier temperature and type, etc)
affects the measurement by introducing  an \textit{additional} noise which is
independent of the sample state, so by taking the difference between
the two noise powers one can get rid of the instrumentation-dependent noise power.

 Equilibrium properties, can not be described by the excess noise
since by definition it vanishes in equilibrium.

Since in most cases mesoscopic samples are driven out of equilibrium by
 an external DC voltage, $V$, we shall consider the quantity:
\begin{eqnarray}\label{excess noise V defined}
S_{M,excess}(\Omega)= S_{M,V}(\Omega) - S_{M,0}(\Omega),
\end{eqnarray}
however, other means (e.g., by application of external radiation \cite{Prober photons},\cite{Glattli photons}) can be used
 to drive the system out of equilibrium.

\subsection{Statement of the problem}
\label{sec statement of problem}

What quantity  should one calculate  in order to predict $S_{M,excess}(\Omega)$?
Will this quantity depend on the properties of the sample only,
or also on the particular experimental setup?

To answer these questions we first consider the classical case.

\section{Which quantity is measured in a classical noise measurement?}
\label{sec which quant is meas Class}

\subsection{The classical case without amplification}
\label{sec Class meas no ampl}

Consider a current $I(t)$ flowing in a system which is in a stationary state.
  Consider a long time interval $T_0\gg \omega $
and define the \textit{restricted fourier transform}
of $I(t)$,
\begin{eqnarray}\label{FT current operator }
I_{T_0}(\omega)=
  \int_{-\frac{T_0}{2}}^{\frac{T_0}{2}}dt
e^{i\omega t}I(t),
\end{eqnarray}
the \textit{power spectrum} of $I(t)$,
\begin{eqnarray}\label{Sw defined classical case}
S_I(\omega)=\lim _{T_0 \rightarrow \infty}\frac{|I_{T_0}(\omega)|^2}{T_0}.
\end{eqnarray}
and the \textit{correlator} of $I(t)$,
\begin{eqnarray}\label{c tau defined }
c(\tau)=c(-\tau)=\overline{I(0)I(\tau)}.
\end{eqnarray}
The Wiener-Khinchin theorem \cite{gardiner stoch handbook} states that
\begin{eqnarray}\label{Weiner Khinchine}
S(\omega)=S(-\omega)=\int_{-\infty}^{\infty}d\tau e^{-i\omega\tau} c(\tau),
\end{eqnarray}
\begin{eqnarray}\label{Weiner Khinchine inversed}
c(\tau)=\frac{1}{\pi}\int_0^{\infty}d\omega cos\omega\tau S_I(\omega).
\end{eqnarray}
and specifically also that,
\begin{eqnarray}\label{C0=int Sw}
c(0)=\overline{I^2(t)}=\frac{1}{\pi}\int_0^{\infty}d\omega S_I(\omega).
\end{eqnarray}

By its definition, a filter greatly reduces the fourier components at frequencies further than
a band width $\Delta_f$ away from its center
frequency, $\Omega$ .
Therefore, assuming a regular behavior of $J_{T_0}(\omega)$, and a small $\Delta_f$,
 the restricted transform of the current coming out of that filter, $I_{f,T_0}(\omega),$
 is related to that of the incoming current,  $J_{T_0}(\omega)$, by
\begin{eqnarray}\label{IMwT0 rectangular relation to JwT0}
I_{f,T_0}(\omega)=\gamma J_{T_0}(\Omega) ~~~~~|\omega-\Omega|\lesssim\Delta_f, \nonumber\\
I_{f,T_0}(\omega)=0~~~~~~~~~~~~~~|\omega-\Omega|\gtrsim\Delta_f,
\end{eqnarray}
where $\gamma$ is constant. An example of a filter is shown in Fig. 2.
For the moment we do not consider the possibility that the filter adds its own thermal noise to
$I_f(t)$.
Thus, according to their definitions, the power spectrum of $I_f(t)$ and $J(t)$
are related by:
\begin{eqnarray}\label{SMw Sw rectangular relation}
S_{f}(\omega)=\gamma^2S_J(\Omega) ~~~~~|\omega-\Omega|\lesssim \Delta_f, \nonumber\\
S_{f}(\omega)=0~~~~~~~~~~~~~~~|\omega-\Omega|\gtrsim\Delta_f.
\end{eqnarray}

Applying Eq. (\ref{C0=int Sw}) to  $I_f(t)$,
Eq. (\ref{Weiner Khinchine}) to $J(t)$,
 and using Eq. (\ref{SMw Sw rectangular relation}) one gets
\begin{eqnarray}\label{C0=JJ}
\frac{1}{\Delta_f }\overline{I_f^2(t)}=\gamma^2S_J(\Omega)=
\gamma^2\int_{-\infty}^{\infty}d\tau e^{i\Omega\tau}
\overline{J(0)J(\tau)}.
\end{eqnarray}
Eq. (\ref{C0=JJ}) clarifies what might be a confusing feature of a noise measurement with a filter:
 taking the square of the current coming out of the filter at one time
yields information on the correlation in the current in the sample at two different times.

The average energy stored in an RLC filter is \cite{Purcell Berkley}:
\begin{eqnarray}\label{Ef}
\langle E_{f} \rangle = L\langle I_f^2 \rangle.
\end{eqnarray}
For later purposes we note that this equation is valid also in a stationary quantum state as a consequence of the virial theorem \cite{Cohen Tannoudji}.
Since, for the moment, the amplification stage is ignored we have $I_f(t)=I_a(t)$.
Therefore, making use of the definition of the measured noise, Eq. (\ref{SMw defined}), we see that
\begin{eqnarray}\label{what is measured classicaly}
S_M^{\textrm{(no ampl)}}(\Omega)=\frac{1}{\Delta_fL}\langle E_f \rangle=
\gamma^2\int_{-\infty}^{\infty}d\tau e^{i\Omega\tau}
\langle J(0)J(\tau)\rangle.
\end{eqnarray}

\begin{figure}
     \begin{center}
         \includegraphics[height=1.5in,width=4.6in,angle=0]{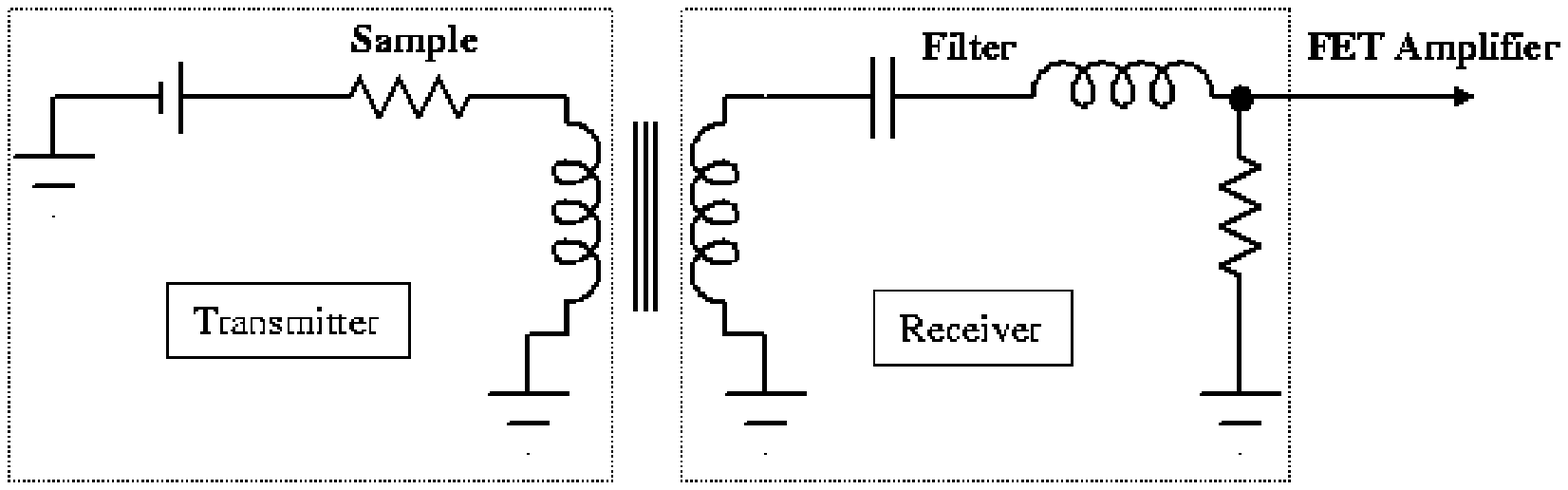}
\vspace{-0.1cm}
         \caption{Inductive coupling to an RLC filter}
     \end{center}
 \end{figure}

\subsection{The classical case with amplification}
\label{sec Class meas with ampl}
We proceed now to include the amplification stage. In order
to avoid complications due to the differences between different types of amplifiers we assume here
that only the cosine component of the output current, $I_{a,c}(t),$ is measured, squared and averaged.
 Such an assumption enables us to assign a single gain, $G$, to the amplifier (whether it is phase
 sensitive or not) that multiplies the incoming signal.
 Thus, $I_f(t)$ enters the amplifier and $I_a(t)$ comes out of it.
 Then, the component $I_{a,c}(t)=GI_{f,c}(t)$
 is measured and squared. But $\overline{I_{f,c}^2}=\overline{I_{f}^2}$ and
 therefore by Eqs. (\ref{C0=JJ})-(\ref{what is measured classicaly}) one has:
\begin{eqnarray}\label{what is measured classicaly with ampl}
S_M(\Omega)=
 \frac{G^2}{\Delta_fL}\langle E_f \rangle=
\bar{G}^2\int_{-\infty}^{\infty}d\tau e^{i\Omega\tau}
\langle J(0)J(\tau)\rangle.
\end{eqnarray}
where $\bar{G}=\gamma G.$
\subsection{Thermal noise produced by the setup}
\label{sec thermal noise class}
Even when considered classically, the measurement setup can perform
  according to Eq. (\ref{Lin ampl defined}) and (\ref{what is measured classicaly with ampl})
  only as long as its components are operating at temperatures which are low
  compared with the signal power spectrum.
At higher temperatures, one should take into account the thermal noise, $S_{N,T}(\omega),$ produced by these components
 and add it to the output signal.
We shall not discuss the particular form of this noise, except for
 mentioning that in equilibrium and at low frequencies
its contribution is $k_BT$ (the Nyquist-Johnson noise) times the amplifier-gain squared, and that it is independent
of the  input, i.e., of the state of the sample.
Thus, we write:
\begin{eqnarray}\label{what is measured class+ampl + therm noise}
S_M(\Omega)=
 \frac{G^2}{\Delta_fL}\langle E_f \rangle +S_{N,T}(\Omega)=
G^2\int_{-\infty}^{\infty}d\tau e^{i\Omega\tau}
\langle J(0)J(\tau)\rangle+S_{N,T}(\Omega)\nonumber \\
\end{eqnarray}
where $S_{N,T}(\Omega)$ is defined as the noise measured with no input.

\noindent Comments:

1. Other types of noise such (e.g., $1/f$ noise)
 may occur inside the setup components.
 However, unlike the noise required by thermodynamics (and in the quantum case also that
 required by the Heisenberg principle), these may in principle
 be eliminated and thus are not considered here.

 2. Adding a setup noise which is assumed to be independent of the input,
 is justified by assuming that the statistical
 distributions of the state of the total system is a product of that of the amplifier (and the detector) and the
 filter+sample and that the coupling between those parts is weak. In the
 quantum case the assumption is that the density matrix of the system is a product of that
 of the amplifier (and the detector) and the sample+filter. For more details see the appendix.

\subsection{Classical excess noise}
\label{sec Classical excess noise}

According to Eqs. (\ref{excess noise V defined}), (\ref{what is measured class+ampl + therm noise}),
 the  excess noise in a classical measurement is:
\begin{eqnarray}\label{measured excess classicaly + therm noise}
S_{M,excess}(\Omega)=
 \frac{G^2}{\Delta_fL}\langle E_f \rangle_{excess}=
\bar{G}^2\int_{-\infty}^{\infty}d\tau e^{i\Omega\tau}
\langle J(0)J(\tau)\rangle_{excess}
\end{eqnarray}
where
\begin{eqnarray}\label{excess filter energy class}
\langle E_f \rangle_{excess}=
\langle E_f \rangle_V - \langle E_f \rangle_0,
\end{eqnarray}
and
\begin{eqnarray}\label{excess correlt class}
\langle J(0)J(\tau)\rangle_{excess}=
\langle J(0)J(\tau)\rangle_{V}-\langle J(0)J(\tau)\rangle_{0},
\end{eqnarray}
are the differences in the filter energies and the correlators in and out of equilibrium.
Eqs. (\ref{what is measured class+ampl + therm noise}) and (\ref{measured excess classicaly + therm noise})
tell us what are the measured quantities in noise and excess noise measurements in
classical situations, i.e., when quantum effects can be neglected. They show that although
the measured noise, Eq. (\ref{what is measured class+ampl + therm noise}), is
setup-dependent because the term $S_{N,T}(\Omega)$ depends on the setup type
and temperature, the measured excess noise is not.

Eq. (\ref{measured excess classicaly + therm noise}) also gives a simple physical
picture to the excess noise: when a voltage is applied to the sample and drives it out of equilibrium,
 the current (or charge) fluctuations in the sample change (typically, they increase),
and interact with the charges or currents in the RLC circuit (e.g., through capacitive or inductive coupling)
causing an increase in the  energy flow from the sample into the circuit in a similar way to that
in which the current fluctuations in an antenna of a classical transmitter emit energy into a receiver (Fig. 2).
Part of this energy is accumulated in the capacitor and the inductor and part is dissipated in the resistor.
Eventually the system arrives at a stationary (though not an equilibrium) state where the filter
energy is higher than before.
The measured excess noise is  simply this increase in the filter energy multiplied by the amplifier gain.

Having obtained a detailed picture of the classical noise measurement we are now ready to
analyze the quantum case.

\section{Which quantity is measured in quantum noise measurement?}
\label{sec which quant is meas Quant}

When the measured frequency is higher than the temperature of the sample or the
setup, quantum effects become important. One may then consider replacing the
 current $J(t),$ and the average over realizations $\langle J(0)J(\tau)\rangle,$
 in Eqs. (\ref{what is measured class+ampl + therm noise}) and (\ref{measured excess classicaly + therm noise})
  by, respectively, the Heisenberg current operator of the electrons in the sample, $\hat{J}(t),$
   and the  expectation value of the product of operators, $\hat{J}(0)\hat{J}(\tau),$ in the quantum stationary state of the system.
However, in attempting to do so one immediately encounters the following two  questions (which are answered below):

1. The current operator does not commute with itself at different
 times (the product $\hat{J}(0)\hat{J}(\tau)$ is not Hermitian) and therefore
it is not clear in which order, $\hat{J}(0)\hat{J}(\tau)$ or $\hat{J}(\tau)\hat{J}(0)$
the product should be written or whether it should be replaced
by its  \textit{symmetrized} version $(\hat{J}(0)\hat{J}(\tau)+\hat{J}(\tau)\hat{J}(0))/2,$ (which is Hermitian)
 as is customarily suggested in text books \cite{Landau Lifshitz}.

2. What are the properties of the setup noise (the analog $S_{N,T}(\Omega)$) in the quantum case, and specifically,
does it vanish in the limit of zero temperature as was the case in the classical regime?

\subsection{The quantum case without amplification}
\label{sec Quant meas no ampl}
Consider first a mesoscopic system, e.g., a ballistic quantum point contact
between two ohmic contacts,
in which a DC voltage is applied to the left contact and the current
flowing out of the right one interacts with the current
in an RLC circuit (modelled by an harmonic oscillator with a small damping)
through an inductive coupling of the form (see Fig. 2):
   $$\alpha \hat{J}(t)\hat{I}_f(t).$$
The above system
was considered in Refs. \cite{LesovikLoosen} and \cite{Gavish Levinson Imry}  (in the limit of small $\alpha$ and $\Delta_f$).
It was shown,  that
as a result of this interaction there is an energy flow between the electronic system
and the filter. The current fluctuations in the electronic system
excite the harmonic modes of the filter in a similar  way to that in which
 current fluctuations in an antenna excite the photon modes in the electromagnetic field of the vacuum .
 As a result of switching  the \emph{interaction} on adiabatically while keeping the DC voltage constant,
 (unlike what is considered above  where the \textit{voltage} is switched on),
 the energy of the filter was found to increase by an amount of \cite{LesovikLoosen}-\cite{Gavish Levinson Imry}:
 \begin{eqnarray}\label{Delta Ef LC}
\delta \langle E_f \rangle =L(\langle \hat{I}^2_f(t)\rangle_{\alpha}-\langle \hat{I}^2_f(t)\rangle_{\alpha=0})=~~~~~~~~~~~~~
~~~~~~~~~~~~~~~~~~~~~\nonumber \\
 \gamma^2 \left[ (N+1)S_J(\Omega)-NS_J(-\Omega)\right]=
 \gamma^2 \left[ S_J(\Omega)-2 N \hbar \Omega G_d(\Omega)\right],
\end{eqnarray}
where $S_J(\Omega)$ is the transform of the \textit{nonsymmetrized}
quantum correlator,
 \begin{eqnarray}\label{Sw quantum defined non sym}
 S_J(\Omega)=\int_{-\infty}^{\infty}d\tau e^{i\Omega\tau}
\langle \hat{J}(0)\hat{J}(\tau)\rangle,
\end{eqnarray}
where $N$ is the average number of quanta in the oscillator,  and $G_d(\Omega)$ is the differential conductance \cite{endnote fdt}.
$\gamma^2$ is equal to $\alpha^2$ times some multiplicative factors (which anyhow cancel in the setup calibration).
$S_J(\Omega)$ has a physical meaning: it is proportional to the fermi-golden rule emission-rate
of quanta of $\hbar \Omega$  from the sample into the filter \cite{Gavish Levinson Imry}.
Similarly, $S_J(-\Omega)$ is proportional to the absorption rate.
Thus, Eq. (\ref{Delta Ef LC}) has a simple physical meaning: the change in the filter energy
correspond to the spontaneous emission from the sample plus the net energy flow due
to the difference between the induced emission and absorption.

In order to obtain the change in the filter energy due to the application of the voltage,
one has to calculate the difference between finite and zero voltage:
  $\langle E_f \rangle_{excess}=\delta \langle E_f \rangle_V - \delta \langle E_f \rangle_0$. One obtains:
\begin{eqnarray}\label{Ef excess quantum}
 \langle E_f \rangle_{excess}=\gamma^2\int_{-\infty}^{\infty}d\tau e^{i\Omega\tau}
\langle \hat{J}(0)\hat{J}(\tau)\rangle_{excess},
\end{eqnarray}
where it was assumed that the voltage-dependence of the differential conductance is weak.
Eq. (\ref{Ef excess quantum}) expresses the change in the filter energy when the sample
is in and out of equilibrium.  We are now ready to proceed to the final step and take
the amplification into account.

\subsection{The quantum case with amplification}
\label{sec Quant meas with ampl}
Consider the current $\hat{I}_f(t)$ in an RLC circuit which serves as the input signal of
a linear amplifier. The quantum theory of linear amplifiers \cite{Yurke denker}
specifies what limitations  the Heisenberg principle puts on their performances.
 The limitations relevant to our case can be summarized as follows:

An amplifier that amplifies both
sine and cosine components of the input signal \textit{must} add noise,
which we will denote by $S_{N,Q}(\Omega)$, to the measured signal,
and this noise does not vanish at zero temperature.
Therefore, a phase insensitive amplifier must
add noise to the measured signal. However, the added noise is not necessarily
distributed evenly between the two components. The amount of noise which is added,
say, to the cosine component, depends on the particular amplification setup and temperature.
The fluctuation of the current coming out of a phase insensitive amplifier
are given by the form \cite{Yurke denker}:
\begin{eqnarray}\label{Ia2}
 S_M(\Omega)=\Delta_f^{-1}\langle \hat{I}_a^2(t) \rangle=
G^2\Delta_f^{-1}\langle \hat{I}_f^2(t) \rangle +S_{N,Q}(\Omega)
\end{eqnarray}
where $S_{N,Q}(\Omega)$ depends only on the properties and state of the amplifier and detector while
$\langle \hat{I}_f^2(t) \rangle$ depends only on those of the filter.
Similarly, the fluctuation of the current coming out of a phase sensitive amplifier (which
 can be, ideally, noiseless and which amplifies only the cosine component and disposes of the sine component)
are given by the form:
\begin{eqnarray}\label{Iac2}
  S_M(\Omega)=\Delta_f^{-1}\langle \hat{I}_{a,c}^2(t) \rangle=G^2\Delta_f^{-1}\langle \hat{I}_{f,c}^2(t) \rangle.
\end{eqnarray}
A brief  review on the origin of the additive form of the right hand side of Eq. (\ref{Ia2}) is given in the appendix.
\section{Excess quantum noise measurement}
\label{sec excess quan noise meas}
Since the filter is in a stationary state,
one has $\langle \hat{I}_f^2\rangle = \langle \hat{I}_{f,c}^2 \rangle
=\frac{1}{L}\langle E_f\rangle$ and therefore, Eqs. (\ref{Ef}), (\ref{Ia2}) and (\ref{Iac2}) yield:
\begin{eqnarray}\label{SM quantum}
  S_{M,excess}(\Omega)=G^2(L\Delta_f)^{-1}\langle E_f \rangle_{excess}.
\end{eqnarray}

Eqs. (\ref{Ef excess quantum}) and (\ref{SM quantum}) imply
\begin{eqnarray}\label{SM excess quantum}
  S_{M,excess}(\Omega)=\bar{G}^2\int_{-\infty}^{\infty}d\tau e^{i\Omega\tau}
\langle \hat{J}(0)\hat{J}(\tau)\rangle_{excess},
\end{eqnarray}
where $\bar{G}=\gamma G.$

Eq. (\ref{SM excess quantum}) is our main result.
It shows that in order to predict the result of an excess noise measurement
one should calculate the correlators (no symmetrization is required) in and out of equilibrium
and take the difference between them.
It also shows that there will be no contribution from the zero-point
 fluctuations since $S(\Omega>0)$
(the \emph{emission} spectrum)   does not contain such contribution - the zero point
fluctuations can not emit energy.

When the $sample$ (but not necessarily the setup) is at zero temperature the
correlator vanishes in equilibrium since the sample can not emit anything.
Therefore:
\begin{eqnarray}\label{SM excess quantum T=0}
  S_{M,excess}(\Omega)=G^2\int_{-\infty}^{\infty}d\tau e^{i\Omega\tau}
\langle \hat{J}(0)\hat{J}(\tau)\rangle, ~~~~~~~~~k_BT_s=0.
\end{eqnarray}

We note that the equality in Eq. (\ref{SM excess quantum}) is not term-by-term.
The excess  noise is equal to the excess of the correlator
but the noise by itself is not equal to the correlator by itself since according to Eq. (\ref{Ia2})
the former depends also on the setup properties
  while the latter depends only on the sample properties.

The derivation of Eq. (\ref{SM excess quantum}) is valid for all
linear amplifiers used in noise measurements in mesoscopic systems
and the parametric ones which are analyzed in Ref. [4]. The
condition that the conductance $G_d$ remains constant when the DC
voltage is turned on, that was used in deriving this equation, may
be understood by considering the zero temperature case: the excess
noise is the power flow of energy from the sample into the filter.
In order to enable an efficient measurement of this power an
impedance matching is needed between the sample, the transmission
lines and the filter. If the sample conductance is very different
in its equilibrium and nonequilibrium states, then, initially good
impedance-matching  with the detector that enables an efficient
power flow in equilibrium, means a bad impedance matching out of
equilibrium with an inefficient power flow. In such a case one
should correct Eq. (\ref{SM excess quantum}) by taking into
account the different impedance ratios in and out of equilibrium.

\section{Suggested verifications of the theory}
\label{sec verification}
A straightforward way to verify Eq. (\ref{SM excess quantum T=0}) is to measure the excess noise
 in a single-channel ballistic quantum point contact at high frequencies,  $\hbar\Omega \gtrsim eV.$
For such a system the nonsymmetrized correlator is given by (see
Ref. [14] for the symmetrized version, and [15] and [13] for the
nonsymmetrized one):
\begin{eqnarray}\label{S balist nonsym}
  S(\Omega,T_{s},V)=
  \frac{e^2}{h}|t|^2(1-|t|^2)\sum_{\epsilon=\pm 1}F(\hbar \Omega+\epsilon eV)+\frac{e^2}{h}|t|^4F(\hbar\Omega)
\end{eqnarray}
where $F(x)=x(e^{x/k_BT_{s}}-1)^{-1}$, $T_s$ is the sample temperature and
 $|t|^2$ is the transmission of the channel.
According to our theory, such an excess noise measurement would yield
$S(\Omega,T_{s},V)-S(\Omega,T_{s},0)$  for any amplifier type or temperature,
while without taking the excess the result will generally depend
substantially on type of the setup and  its temperature. In particular, for $T_s \ll eV,\hbar\Omega$,
 the excess noise measurement will yield $S(\Omega,0,V)$, i.e.,
the nonsymmetrized correlator (with no contribution from the zero point fluctuations)
while without taking the excess the result will generally differ from both the non-symmetrized
and the symmetrized correlators and will depend on the particular setup.\\

\section{Appendix. Noise added in amplification}
\label{appendix}

\subsection{Requirements from quantum linear amplifier output}

Consider an RLC circuit (see e.g., Fig. 2), which we shall call 'the input', connected into a linear amplifier.
 Let $\hat{I}_{f}(t)$ be the Heisenberg operator of the current
in the input. This operator acts on the degrees of freedom of the circuit and therefore its expectation values
are determined when the circuit state is given. Let $\hat{I}_{a}(t)$ be the Heisenberg operator of the current at
the output port of the amplifier. In general, this operator  acts on both the input and the amplifier degrees of freedom.
For an ideal linear amplifier:
\begin{eqnarray}
\label{Lin ampl ideal quantum}
\hat{I}_a(t)= \hat{I}_{a,c}(t)\cos\Omega t
+\hat{I}_{a,s}(t)\sin \Omega t=G_1 \hat{I}_{f,c}(t)\cos\Omega t
+G_2\hat{I}_{f,s}(t)\sin \Omega t.
\end{eqnarray}
Note that in such an ideal case the output operator, $\hat{I}_a(t)$, acts only on the degrees of freedom of the input (and not
any of the setup) - the input state determines completely the output independently of the amplifier state.
However, the Heisenberg principle limits on the possibility of realizing such a device:
it requires the sine and the cosine components to obey a minimum uncertainty
 relation (see Eq. (2.19) in Ref. [4]).
Suppose  one constructs a minimum-uncertainty state for $\hat{I}_{a,c}(t)$ and $\hat{I}_{a,s}(t),$ so that
they can be measured simultaneously at the maximum accuracy permitted without violating the Heisenberg principle.
If $G_1G_2> 1$ (as is the case for a phase insensitive amplifier where $G_1=G_2 \gg 1$)
 then $\hat{I}_{f,c}(t)$ and $\hat{I}_{f,s}(t)$  could have been measured simultaneously
up to an accuracy which is  $G_1G_2$ times better than allowed simply by measuring
$\hat{I}_{a,c}(t)$ and $\hat{I}_{a,s}(t),$ and then applying Eq. (\ref{Lin ampl ideal quantum}).
Therefore, an amplifier with $G_1G_2> 1$ and specifically a phase insensitive
amplifier is forbidden in quantum mechanics.

One way to overcome these limitations is to build an amplifier in which, for example, $G_1\gg 1, ~  G_1 G_2=1,$
that is, a phase sensitive amplifier that amplifies the cosine component and diminishes the sine component so that in the limit
of $G_1=G \rightarrow \infty$ we can write:
\begin{eqnarray}
\label{Lin ampl ph sensitive  quantum}
\hat{I}_a(t)= G \hat{I}_{f,c}(t)\cos\Omega t
\end{eqnarray}
An example for such a device is the degenerate parametric amplifier \cite{Yurke denker}.
Another way to overcome the above limitations is to allow $G_1 G_2>1,$ (and in particular  $G_1 =G_2=G\gg 1,$)
but add to the right hand side of Eq. (\ref{Lin ampl ideal quantum}) an additional term that will
operates on the amplifier degrees of freedom so that:
\begin{eqnarray}
\label{Lin ampl with noise term quantum}
\hat{I}_a(t)=G\hat{I}_f(t) + \hat{I}_{N,Q}(t),
\end{eqnarray}
where $\hat{I}_{N,Q}(t)$ is an operator acting on the amplifier degrees of freedom such
as the electronic state in a field effect transistor or the idler resistor state in
a non-degenerate parametric amplifier.

\subsection{Independence of the amplifier noise of the sample state}
Let us assume that the input is in a stationary state in which the average currents vanish and take the expectation square
of the output current, as is done in a noise measurement.
In order to do so we should specify the state of the system or more generally, the density matrix.
Here we make an important assumption that the total density matrix of the system, $\rho_s$ is a product
of the amplifier density matrix, $\rho_a,$ and the input one, $\rho_f$: $\rho_a$ and $\rho_f$,  $\rho_s=\rho_f\rho_a.$
Such an assumption is justified, e.g., when the interaction between the amplifier and the input is small compared to their coupling
with the thermal baths that determine their temperatures.
Then using Eq. (\ref{Lin ampl ph sensitive  quantum})
and averaging over time
much larger than $\Omega^{-1}$ one has
\begin{eqnarray}
\label{noise in ph sensitive quantum}
\langle \hat{I}^2_a(t)\rangle_s= \frac{1}{2}G^2\langle \hat{I}^2_{f,c}(t)\rangle_{f,a}
= \frac{1}{2}G^2 \langle \hat{I}^2_{f,c}(t)\rangle_f ,
\end{eqnarray}
where $\langle A \rangle_x = Tr\rho_xA$.
This relationship is
identical to Eq. (\ref{Iac2}) except for the factor 1/2 which appeared due to different
definitions of the gain. In the phase insensitive case one gets from Eq. (\ref{Lin ampl with noise term quantum})
\begin{eqnarray}
\label{noise in ph insensitive quantum}
\langle\hat{I}^2_a(t)\rangle_s
=G^2\left[ \langle \hat{I}^2_f(t)\rangle_f +\langle \hat{I}^2_f(t)\rangle_a
+2\langle \hat{I}_f(t)\rangle_f \langle \hat{I}_a(t)\rangle_a \right]\nonumber \\
=G^2 \langle  \hat{I}^2_f(t)\rangle_f +S_{N,Q}(\Omega)\Delta_f~~~~~~~~~~~~~~~~~~~~~~~~
\end{eqnarray}
where $S_{N,Q}(\Omega)\Delta_f=G^2\langle  \hat{I}^2_f(t)\rangle_a$, as in Eq. (\ref{Ia2}).
Eqs. (\ref{noise in ph sensitive quantum}) and (\ref{noise in ph insensitive quantum}) shows that the amplifier
noise is \emph{additive} to the input noise (in Eq. (\ref{noise in ph sensitive quantum}) it is trivially zero),
at least as long as one assumes that the density matrix of the system can be factorized
as described above.

\section{Acknowledgment}
This project was supported by the Israel Science Foundation, by
the German-Israeli Foundation (GIF), Jerusalem, and by the Maurice
and Gabriella Goldschleger Center for Nanophysics at the Weizmann
Institute. E. Comforti, R. dePiccioto, C. Glattli, M. Heiblum, W.
Oliver, D. Prober, M. Reznikov, U. Sivan and A. Yacoby are warmly
thanked for many discussions, and for providing detailed
descriptions of their experimental setups. U. G. would like to
express special thanks to B. Doucot for his support and for many
very helpful discussions.

\end{document}